\documentclass[letterpaper,journal]{IEEEtran}

\usepackage{fix-cm}
\usepackage{amsmath,amssymb,amsfonts,mathtools,empheq}
\usepackage{array,booktabs}
\usepackage{graphicx}
\usepackage{cite}

\DeclareMathSizes{9.8}{9.8}{6.8}{4.8}

\newcommand{\mathoneptsmaller}[1]{%
  \mathchoice
    {\text{\small$\displaystyle #1$}}
    {\text{\small$\textstyle #1$}}
    {\text{\small$\scriptstyle #1$}}
    {\text{\small$\scriptscriptstyle #1$}}%
}

\newcommand{\halfptsmallersum}[1]{%
  \mathop{\scalebox{0.95}{$\displaystyle\sum$}}%
  \limits_{\scalebox{0.928571}{$\scriptstyle #1$}}%
}

\newtheorem{definition}{Definition}
\newtheorem{proposition}{Proposition}
\renewenvironment{definition}[1][]{%
  \refstepcounter{definition}%
  \par\noindent
  \textbf{Definition~\thedefinition%
    \if\relax\detokenize{#1}\relax\else\ (#1)\fi:}\enspace
  \ignorespaces
}{\par}

\begin{document}

\title{A Counterexample to Two Representative Unit Aggregation Formulations for Unit Commitment}

\author{Zixuan Duan, \IEEEmembership{Student Member, IEEE}, and
Zhengshuo Li, \IEEEmembership{Senior Member, IEEE}%
\thanks{This work was supported by the National Natural Science Foundation of China under Grant 52377107,
the Shandong Provincial Natural Science Foundation under Project ZR2026QB17, and the Taishan Scholars Program.}%
\thanks{Zixuan Duan and Zhengshuo Li are with Shandong University, Jinan, China. 
Zhengshuo Li is the corresponding author (email: zsli@sdu.edu.cn).}}

\maketitle

\begin{abstract}
Unit aggregation removes symmetry among identical generators in unit commitment, 
but an aggregate formulation is feasible-region exact only if every aggregate trajectory 
it admits has a feasible unit-level realization. This letter shows that two commonly used aggregation formulations 
for slow-ramping units, i.e., \textit{p-clustered unit commitment} (PCUC) and \textit{tight unit aggregation} (TUA), 
cannot satisfy this requirement.
We construct a counterexample that satisfies all aggregate constraints of both formulations 
yet admits no feasible disaggregation. 
This counterexample reveals a limitation common to both formulations: 
they do not guarantee intertemporal consistency of unit-level output allocations.
Experiments on the replication cases reported in published literature 
further verify that such infeasibility of disaggregation can even occur in optimal solutions of aggregate models.
This letter demonstrates that PCUC and TUA can still overestimate ramping flexibility. 
Developing exact aggregation models that can be solved efficiently for identical slow-ramping units remains a challenge.
\end{abstract}

\vspace{-1pt}
\begingroup
% Reduce all nomenclature text and headings from 10pt to 9.8pt.
\renewcommand{\normalsize}{\fontsize{9.8pt}{12pt}\selectfont}
\normalsize
\section*{Nomenclature}
\addcontentsline{toc}{section}{Nomenclature}
\vspace{-2pt}
\subsection*{A. Sets and Indices}
\begin{IEEEdescription}[\IEEEusemathlabelsep\IEEEsetlabelwidth{$\mathcal G_k$}]
\item[$g / \mathcal G$] Index/set of generating units.
\item[$\mathcal G^1$] Set $\{g\in\mathcal G:TU_g=1\}$.
\item[$k / \mathcal K$] Index/set of identical-unit clusters.
\item[$\mathcal K^1$] Set $\{k\in\mathcal K:TU_k=1\}$.
\item[$\mathcal G_k$] Set of generating units in cluster $k$.
\item[$t / \mathcal T$] Index/set of periods, with $\mathcal T=\{1,\ldots,T\}$.
\end{IEEEdescription}
\vspace{-3.5pt}
\subsection*{B. Parameters}
\begin{IEEEdescription}[\IEEEusemathlabelsep\IEEEsetlabelwidth{$SUSD$}]
\item[$\underline P/\overline P$] Minimum/maximum output.
\item[$RU/RD$] Normal ramp-up/ramp-down limit.
\item[$SU/SD$] Startup/shutdown physical-output capability.
\item[$TU/TD$] Minimum up/down time.
\end{IEEEdescription}
\vspace{-3.5pt}
\subsection*{C. Variables}
\begin{IEEEdescription}[\IEEEusemathlabelsep\IEEEsetlabelwidth{$W$}]
\item[$u_{gt}$] On/off binary variable of unit $g$ at period $t$.
\item[$v_{gt}$] Startup binary variable of unit $g$ at period $t$.
\item[$w_{gt}$] Shutdown binary variable of unit $g$ at period $t$.
\item[$p_{gt}$] Above-minimum output of unit $g$ at period $t$.
\item[$U_{kt}$] Number of online units in cluster $k$ at period $t$.
\item[$V_{kt}$] Number of starting units in cluster $k$ at period $t$.
\item[$W_{kt}$] Number of shutting-down units in cluster $k$ at period $t$.
\item[$P_{kt}$] Above-minimum output of cluster $k$ at period $t$.
\end{IEEEdescription}
\endgroup
\section{Introduction}
\label{sec:introduction}
\IEEEPARstart{U}{nit} commitment (UC) models often contain multiple generating units with identical operating parameters \cite{DuiChenReview}.
Permuting the scheduling and power outputs of these units leaves the total cost unchanged,
potentially causing the branch-and-bound (B\&B) algorithm to spend considerable time
exploring redundant nodes.
In mixed-integer linear programming (MILP), this interchangeability 
is known as symmetry \cite{IntegerProgrammingSymmetry}. 
Unit aggregation provides an effective means of eliminating such symmetry. 
Previous studies \cite{AggregationEF} have established that the clustered unit commitment (CUC) 
formulations presented in \cite{AggregationLink} are exact for \textit{fast-ramping} units. 
Here, fast-ramping units refer to generators whose ramping capabilities are sufficiently high to cover 
the relevant operating range within a single scheduling interval. The aggregation of the remaining units, 
referred to in this paper as \textit{slow-ramping} units, has continued to attract considerable research attention 
in recent years. Representative formulations include extended formulation (EF)~\cite{AggregationEF}, 
p-clustered unit commitment (PCUC)~\cite{AggregationLinkIndiv}, and tight unit aggregation
(TUA)~\cite{AggregationTUA}. These formulations aim to reduce the computational burden caused by
symmetry in UC models. EF achieves exact aggregation, 
but its large model size can make it computationally demanding \cite{AggregationEF,AggregationLinkIndiv}.

The central contribution of this letter is to demonstrate that the two aggregation
formulations proposed in \cite{AggregationTUA,AggregationLinkIndiv} remain flawed.
We construct a counterexample that satisfies all aggregate constraints
of both formulations but admits no feasible unit-level realization. 
In other words, both aggregate formulations contain
\textit{spurious} feasible points. 
This counterexample reveals a common limitation: neither formulation
guarantees consistent unit-level output allocations over time.
These results show that PCUC and TUA can still overestimate
ramping flexibility.
Together with the computational burden of EF, these limitations
highlight that developing efficiently solvable exact aggregation models remains a challenge.
\section{Preliminaries and Existing Aggregation Formulations}
\label{sec:preliminaries}
\subsection{Unit-Level Unit Commitment Formulation}
\label{subsec:unit-level-model}
The UC problem minimizes the total generation cost over
the scheduling periods:
\begin{equation}
\mathrm{Minimize}\quad
\textit{operating costs}+\textit{startup costs}.
\label{obj}
\end{equation}
The system constraints are as follows:
{\small
\begin{equation}
\textit{\fontsize{9.3pt}{11pt}\selectfont system-wide constraints, e.g., power balance, branch capacity}
\label{SystemConstr1}
\end{equation}
}%
The complete objective function and system constraints are provided in our online document \cite{ReplicationCodeData}.
The unit operating constraints are formulated as follows:

\allowdisplaybreaks[4]
{\small
\begin{empheq}[left=\empheqlbrace]{gather}
u_{gt}-u_{g,t-1}=v_{gt}-w_{gt}\quad\forall g,t
\label{UnitConstr1}\\
\sum_{i=t-TU_{g}+1}^{t}v_{gi}\leq u_{gt}\quad\forall g,t
\label{UnitConstr2}\\
\sum_{i=t-TD_{g}+1}^{t}w_{gi}\leq 1-u_{gt}\quad\forall g,t
\label{UnitConstr3}\\
\begin{aligned}
p_{gt}\leq{}(\overline{P}_{g}-\underline{P}_{g})u_{gt}-&(\overline{P}_{g}-SU_{g})v_{gt}
\quad\forall g\in\mathcal{G}^{1},t
\end{aligned}
\label{UnitConstr4}\\
\begin{aligned}
p_{gt}\leq{}(\overline{P}_{g}-\underline{P}_{g})u_{gt}
-(\overline{P}_{g}-SD_{g})w_{g,t+1}
\quad\forall g\in\mathcal{G}^{1},t
\end{aligned}
\label{UnitConstr5}\\
\begin{aligned}
p_{gt}\leq{}&(\overline{P}_{g}-\underline{P}_{g})u_{gt}
-(\overline{P}_{g}-SU_{g})v_{gt}\\
&-(\overline{P}_{g}-SD_{g})w_{g,t+1}
\quad\forall g\notin\mathcal{G}^{1},t
\end{aligned}
\label{UnitConstr6}\\
\begin{aligned}
p_{gt}-p_{g,t-1}\leq{}RU_{g}u_{g,t-1}
+(SU_{g}-\underline{P}_{g})v_{gt}
\quad\forall g,t
\end{aligned}
\label{UnitConstr7}\\
\begin{aligned}
-p_{gt}+p_{g,t-1}\leq{}RD_{g}u_{gt}
+(SD_{g}-\underline{P}_{g})w_{gt}
\quad\forall g,t
\end{aligned}
\label{UnitConstr8}
\end{empheq}
}%
Constraint~\eqref{UnitConstr1} links changes in the commitment status to startup and shutdown decisions, while 
constraints~\eqref{UnitConstr2} and \eqref{UnitConstr3} enforce the minimum up- and down-time requirements, respectively. 
Constraints~\eqref{UnitConstr4}-\eqref{UnitConstr6} bound the permissible output range of the units. 
Constraints~\eqref{UnitConstr7} and \eqref{UnitConstr8} enforce the ramp-rate limits between consecutive time intervals.

\subsection{CUC Formulation}
\label{subsec:CUC-model}
The CUC formulation in \cite{AggregationLink} replaces unit-level binary commitment variables
with integer counts of online, starting, and shutting-down units and uses aggregate power
variables to represent their output. This aggregation eliminates symmetry and reduces the
number of discrete variables.
For every $k\in\mathcal K$, let $\mathcal G_k\subseteq\mathcal G$ contain the $G_k=|\mathcal G_k|$ 
identical units. Define
\begin{equation}
\scalebox{0.98}{$\displaystyle
(U_{kt},V_{kt},W_{kt},P_{kt})
=\sum\nolimits_{g\in\mathcal G_k}(u_{gt},v_{gt},w_{gt},p_{gt})
\hspace{0.5em} \forall k,t.
$}
\label{eq:aggregate-variables}
\end{equation}
The CUC formulation is given by
{\small
\begin{empheq}[left=\empheqlbrace]{gather}
U_{kt},V_{kt},W_{kt}\in\{0,1,\ldots,G_k\},\quad P_{kt}\ge0,
\quad \forall k,t.
\label{eq:cuc-domains}\\
U_{kt}-U_{k,t-1}=V_{kt}-W_{kt}
\quad \forall k,t.
\label{eq:count-balance}\\
\sum_{i=t-TU_k+1}^t V_{ki}\le U_{kt}
\quad \forall k,t.
\label{eq:count-mu}\\
\sum_{i=t-TD_k+1}^t W_{ki}\le G_k-U_{kt}
\quad \forall k,t.
\label{eq:count-md}\\
\begin{aligned}
P_{kt}\le{}&(\overline P_k-\underline P_k)U_{kt}
-(\overline P_k-SU_k)V_{kt}\\
&-(\overline P_k-SD_k)W_{k,t+1}
\quad \forall k\notin\mathcal K^1,t.
\end{aligned}
\label{eq:cuc-cap-tu2}\\
\begin{aligned}
P_{kt}&\le{}(\overline P_k-\underline P_k)U_{kt}
-(\overline P_k-SD_k)W_{k,t+1}\\
&-\max\{SD_k-SU_k,0\}V_{kt}\quad \forall k\in\mathcal K^1,t.
\end{aligned}
\label{eq:cuc-cap-tu1-su}\\
\begin{aligned}
\quad &P_{kt}\le{}(\overline P_k-\underline P_k)U_{kt}
-(\overline P_k-SU_k)V_{kt}\\
&\quad-\max\{SU_k-SD_k,0\}W_{k,t+1}\hspace{0.5em} \forall k\in\mathcal K^1,t.
\end{aligned}
\label{eq:cuc-cap-tu1-sd}\\
\begin{aligned}
P_{kt}-P_{k,t-1}\le{}RU_kU_{k,t-1}
+(SU_k-\underline P_k)V_{kt}
\quad \forall k,t.
\end{aligned}
\label{eq:cuc-ru}\\
\begin{aligned}
-P_{kt}+P_{k,t-1}\le{}RD_kU_{kt}
+(SD_k-\underline P_k)W_{kt}
\quad \forall k,t.
\end{aligned}
\label{eq:cuc-rd}
\end{empheq}
}%
Constraints~\eqref{eq:count-balance}--\eqref{eq:cuc-rd} are aggregate counterparts of 
the unit-level operating constraints.

\subsection{Feasible-Region Exactness of Unit Aggregation}
\label{subsec:aggregation-exactness}

\begin{definition}[Feasible-region exactness of aggregation]
\label{def:aggregation-exactness}
An aggregate formulation is \emph{feasible-region exact} if it simultaneously
satisfies the following two conditions:
\begin{enumerate}
\item[(i)] \emph{Completeness}: every feasible unit-level schedule has a
feasible aggregate representation.
\item[(ii)] \emph{Unit-Level Realizability}: every point admitted by the
aggregate formulation has a feasible unit-level realization.
\end{enumerate}
\end{definition}

Cluster $k$ is classified as \emph{fast-ramping} if both its normal ramp-up and
ramp-down limits cover the full operating range within one period, and as
\emph{slow-ramping} otherwise. This classification induces the following partition of $\mathcal K$:

{\small
\begin{equation}
\left\{
\begin{aligned}
\mathcal K_F
&:=\{k\in\mathcal K:\min\{RU_k,RD_k\}
\ge \overline P_k-\underline P_k\},\\
\mathcal K_S
&:=\mathcal K\setminus\mathcal K_F.
\end{aligned}
\right.
\label{eq:ramping-class-partition}
\end{equation}
}%
As established in \cite{AggregationEF}, the
CUC formulation is guaranteed to be feasible-region exact only for the
clusters in $\mathcal K_F$.
Subsequent studies have attempted to extend aggregation to $\mathcal K_S$, 
including PCUC~\cite{AggregationLinkIndiv} and TUA~\cite{AggregationTUA}.

\subsection{PCUC Formulation}
\label{subsec:pcuc-model}
PCUC~\cite{AggregationLinkIndiv} augments the CUC with ordered virtual-unit variables. Let
$\widetilde u_{g,k,t}\in\{0,1\}$ and $\widetilde p_{g,k,t}\ge0$ denote,
respectively, the status and above-minimum output of virtual-unit $g$.
Equations~\eqref{eq:cuc-domains}--\eqref{eq:cuc-cap-tu1-sd} and
\eqref{eq:pcuc-slot-order}--\eqref{eq:pcuc-slot-rd} together constitute the PCUC formulation.

{\small
\begin{empheq}[left=\empheqlbrace]{gather}
\widetilde u_{g+1,k,t}\le\widetilde u_{g,k,t}
\quad \forall k,t,\ g=1,\ldots,G_k-1.
\label{eq:pcuc-slot-order}\\
(U_{kt},P_{kt})
=\sum_{g\in\mathcal G_k}
(\widetilde u_{g,k,t},\widetilde p_{g,k,t})
\quad \forall k,t.
\label{eq:pcuc-slot-links}\\
0\le\widetilde p_{g,k,t}\le
(\overline P_k-\underline P_k)\widetilde u_{g,k,t}
\quad \forall k,g,t.
\label{eq:pcuc-slot-cap}\\
\begin{aligned}
\widetilde p_{g,k,t}\le{}(SU_k-\underline P_k)\widetilde u_{g,k,t}
&+(\overline P_k-SU_k)\widetilde u_{g,k,t-1}\\
&\quad \forall k\notin\mathcal K^1,g,t.
\end{aligned}
\label{eq:pcuc-slot-start}\\
\begin{aligned}
\widetilde p_{g,k,t}\le{}(SD_k-\underline P_k)\widetilde u_{g,k,t}
&+(\overline P_k-SD_k)\widetilde u_{g,k,t+1}\\
&\quad \forall k\notin\mathcal K^1,g,t.
\end{aligned}
\label{eq:pcuc-slot-stop}\\
\begin{aligned}
\widetilde p_{g,k,t}\le{}(\overline P_k-SU_k)\widetilde u_{g,k,t-1}+(\overline P_k-SD_k)\widetilde u_{g,k,t+1}\\
+(SU_k-\overline P_k+SD_k-\underline P_k)\widetilde u_{g,k,t}\quad\forall k\in\mathcal K^1,g,t.
\end{aligned}
\label{eq:pcuc-slot-tu1}\\
\widetilde p_{g,k,t}-\widetilde p_{g,k,t-1}
\le RU_k\widetilde u_{g,k,t}
\quad \forall k,g,t.
\label{eq:pcuc-slot-ru}\\
\widetilde p_{g,k,t-1}-\widetilde p_{g,k,t}
\le RD_k\widetilde u_{g,k,t-1}
\quad \forall k,g,t.
\label{eq:pcuc-slot-rd}
\end{empheq}
}%
It is claimed in \cite {AggregationLinkIndiv} that introducing this set of constraints 
into the CUC formulation can accurately represent the ramping flexibility of individual units, 
thereby overcoming the flexibility overestimation problem inherent in the CUC formulation.

\subsection{TUA Formulation}
\label{subsec:tua-model}
TUA partitions $\mathcal K_S$ into $\mathcal K_R$ and $\mathcal K_L$ 
as defined in \cite{AggregationTUA}. It applies a PCUC-inspired treatment to  
$\mathcal K_L$ and deals with $\mathcal K_R$ as follows.
For $k\in\mathcal K_R$, the startup and shutdown ramping horizons are,
respectively,
\begin{equation}
\scalebox{0.99}{$\displaystyle
T_k^+:=\left\lfloor\frac{\overline P_k-SU_k}{RU_k}\right\rfloor,
\quad
T_k^-:=\left\lfloor\frac{\overline P_k-SD_k}{RD_k}\right\rfloor.
$}
\label{eq:tua-ramping-horizons}
\end{equation}
Let $\widehat P_{kt}:=P_{kt}+\underline P_kU_{kt}$ denote total output. 
For $k\in\mathcal K_R$ and $t\in\mathcal T$, the multi-period
output bound is
\begin{equation}
\begin{aligned}
\widehat P_{kt}\mathrel{\mathoneptsmaller{\leq}}\overline P_k U_{kt}
\mathbin{\mathoneptsmaller{-}}
\mathop{\mathoneptsmaller{\sum_{\tau=0}^{\min\{T_k^+,t-1\}}}}
(\overline P_k\mathbin{\mathoneptsmaller{-}}SU_k
\mathbin{\mathoneptsmaller{-}}\tau RU_k)
V_{k,t\mathbin{\mathoneptsmaller{-}}\tau}\\
\mathbin{\mathoneptsmaller{-}}
\mathop{\mathoneptsmaller{\sum_{\tau=0}^{\min\{T_k^-,T-t-1\}}}}
(\overline P_k\mathbin{\mathoneptsmaller{-}}SD_k
\mathbin{\mathoneptsmaller{-}}\tau RD_k)W_{k,t+1+\tau}.
\end{aligned}
\label{eq:tua-multiperiod-cap}
\end{equation}
The ramping constraints in 
\cite{AggregationTUA} are
\begin{align}
\widehat P_{kt}-\widehat P_{k,t-1}
&\le RU_kU_{k,t-1}+SU_kV_{kt},
\label{eq:tua-ramping-hat}\\
\widehat P_{k,t-1}-\widehat P_{kt}
&\le RD_kU_{kt}+SD_kW_{kt}.
\label{eq:tua-ramping-hat-down}
\end{align}
It should be noted that \eqref{eq:tua-ramping-hat} and
\eqref{eq:tua-ramping-hat-down} are not equivalent to \eqref{eq:cuc-ru} and
\eqref{eq:cuc-rd}, respectively. A detailed derivation is provided in our
online document \cite{ReplicationCodeData}. 
For $k\in\mathcal K_R$, 
TUA comprises \eqref{eq:cuc-domains}--\eqref{eq:count-md} and
\eqref{eq:tua-multiperiod-cap}--\eqref{eq:tua-ramping-hat-down}.

Reference~\cite{AggregationTUA} claims that every feasible aggregate solution can be 
disaggregated into a feasible set of unit-level schedules for the identical units in the cluster.
It further claims that an optimal TUA solution can be disaggregated into a feasible 
unit-level solution without loss of optimality.

\section{Counterexamples to Exact Aggregation}
\label{sec:counterexamples}

\subsection{A Common Counterexample to PCUC and TUA}
\label{subsec:pcuc-counterexample}
Consider a 10-period instance with two identical units sharing the following operating parameters:
\begin{equation}
\begin{gathered}
\overline P=40,\quad \underline P=SU=SD=20,\\
RU=RD=10,\quad TU=TD=6.
\end{gathered}
\label{eq:pcuc-params}
\end{equation}
The PCUC \cite{AggregationLinkIndiv} and TUA \cite{AggregationTUA} can yield the aggregate trajectory
\begin{equation}
\begin{aligned}
U&=(1,1,1,2,2,2,1,1,1,0),\\
V&=(1,0,0,1,0,0,0,0,0,0),\\
W&=(0,0,0,0,0,0,1,0,0,1),\\
P&=(0,10,20,20,10,20,20,10,0,0).
\end{aligned}
\label{eq:pcuc-point}
\end{equation}
However, it can be proved that this trajectory has no feasible unit-level realization. 
Consequently, PCUC and TUA violate condition~(ii) of
Definition~\ref{def:aggregation-exactness} and are not feasible-region exact.

\begin{IEEEproof}[Proof sketch]
Let units 1 and 2 denote the physical units that start up at $t=1$ and $t=4$, respectively. Since unit 2
starts up at $t=4$, the allocations of $(p_{1,t},p_{2,t})$ at
$t=4,5$ must be $(20,0)$ and $(10,0)$, respectively. Since unit 1 shuts down at $t=7$, 
$(p_{1,6},p_{2,6})=(0,20)$. Thus, $p_{2,6}-p_{2,5}=20>RU=10$, contradicting the ramp-up
constraint~\eqref{UnitConstr7}. Thus, PCUC and TUA 
overestimate the \textbf{ramping flexibility} of individual units, and
\eqref{eq:pcuc-point} has no feasible unit-level realization.
A complete proof is provided in our
online document \cite{ReplicationCodeData}.
\end{IEEEproof}

Neither PCUC nor TUA guarantees intertemporal consistency in unit-level output allocations.
In this example, the transition from $t=4$ to
$t=5$ requires the allocation of above-minimum output to be
$(p_{1,5},p_{2,5})=(10,0)$, whereas the transition from
$t=5$ to $t=6$, together with unit 1's shutdown at $t=7$, requires it to be
$(p_{1,5},p_{2,5})=(0,10)$. Although both allocations yield the same
aggregate above-minimum output $P_5=10$, no single allocation at $t=5$
is compatible with both transitions.

\subsection{Further Computational Evidence}
\label{subsec:further-cases}
We further evaluated PCUC and TUA on all 20 instances of the Replication
case reported in \cite{AggregationEF,AggregationTUA}. Both formulations
were solved to optimality in every instance, and each resulting aggregate
solution $(U^*,V^*,W^*,P^*)$ was tested for unit-level realizability by solving
\vspace{-2pt}
{
\setlength{\abovedisplayskip}{4pt plus 1pt minus 1pt}
\begin{equation}
\begin{gathered}
\quad\mathllap{\delta^* = {}}\!\min_{u,v,w,p}
\halfptsmallersum{k\in\mathcal K}\halfptsmallersum{t\in\mathcal T}
  \left|\halfptsmallersum{g\in\mathcal G_k}p_{gt}-P_{kt}^*\right|\hspace{-1em} \\[1pt]
\text{s.t.}\hspace{0.5em}
\halfptsmallersum{g\in\mathcal G_k}(u_{gt},v_{gt},w_{gt})
  = (U_{kt}^*,V_{kt}^*,W_{kt}^*),\hspace{0.5em} \forall k,t, \\[-6pt]
(u,v,w,p)\in\mathcal F_{\mathrm{unit}}.
\end{gathered}
\label{eq:disaggregation-test}
\end{equation}
}%
Here, $\mathcal F_{\mathrm{unit}}$ is the feasible set defined by the
unit operating constraints in Section~\ref{subsec:unit-level-model}.
Thus, the aggregate solution admits a feasible unit-level realization
if and only if $\delta^*=0$ (with a tolerance of $10^{-4}$ in our test).
Table~\ref{tab:replication-realizability} reports the optimal objective
values $\delta^*$ of \eqref{eq:disaggregation-test}. The PCUC solutions had no feasible unit-level realization in 5
instances, whereas the TUA solutions had none in all 20. These results further demonstrate that
both formulations can yield optimal aggregate solutions with no feasible
unit-level realization. The test code and data are available in
\cite{ReplicationCodeData}.
\begin{table}
\caption{Disaggregation Test Results ($\delta^*$) for PCUC and TUA}
\label{tab:replication-realizability}
\vspace{-4pt}
\centering
\footnotesize
\setlength{\tabcolsep}{8pt}
\setlength{\aboverulesep}{1.5pt}
\setlength{\belowrulesep}{2pt}
\setlength{\heavyrulewidth}{0.8pt}
\setlength{\lightrulewidth}{0.5pt}
\renewcommand{\arraystretch}{1.1}
\begin{tabular}{@{}c@{\hspace{11pt}}ccc@{\hspace{11pt}}cc@{}}
\toprule
Instance & TUA & PCUC & Instance & TUA & PCUC \\
\midrule
1  & 272.37  & 0          & 11 & 3375.00 & 0    \\
2  & 613.16  & 10.36      & 12 & 6538.35 & 182.35 \\
3  & 615.00  & 0          & 13 & 4965.00 & 33.68 \\
4  & 750.00  & 0          & 14 & 3810.00 & 0    \\
5  & 2060.05 & 0          & 15 & 6860.00 & 269.28 \\
6  & 3145.16 & 0          & 16 & 4875.00 & 0    \\
7  & 3180.00 & 0          & 17 & 5790.00 & 0    \\
8  & 3245.57 & 53.97      & 18 & 4560.00 & 0    \\
9  & 3750.00 & 0          & 19 & 6000.00 & 0    \\
10 & 3520.00 & 0          & 20 & 6750.00 & 0    \\
\bottomrule
\end{tabular}
\par\vspace{-8pt}
\end{table}

\vspace{-6pt}
\section{Conclusion}
\label{sec:conclusion}
This letter provides explicit counterexamples establishing that neither PCUC nor TUA 
is exact. Tests on the Replication case \cite{AggregationEF,AggregationTUA} 
further show that both formulations 
can yield optimal aggregate solutions that admit no feasible unit-level realization.
Therefore, developing exact aggregation models that can be solved efficiently remains a challenge. 
Recent developments can be found in~\cite{AggregationTCUC}.
Future work will investigate exact and efficiently solvable aggregation models.

\bibliographystyle{IEEEtran}

\bibliography{Ref}

@article{DuiChenReview,
  author  = {Zhang, Biyuan and Ding, Tao and Liu, Yuzheng and others},
  title   = {Application of Symmetry-Breaking Techniques in Unit Commitment Problems},
  journal = {Proceedings of the CSEE},
  pages   = {1--16},
  year    = {2026},
  note    = {Early access, in Chinese. Accessed: Aug. 5, 2026. [Online]. Available: \url{https://link.cnki.net/urlid/11.2107.tm.20260121.1740.006}}
}

@book{IntegerProgrammingSymmetry,
  author    = {J{\"u}nger, Michael and Liebling, Thomas M. and others},
  title     = {50 Years of Integer Programming 1958--2008: From the Early Years to the State-of-the-Art},
  publisher = {Springer},
  year      = {2009}
}

@article{AggregationLinkIndiv,
  author    = {Morales-Espa{\~n}a, Germ{\'a}n and Tejada-Arango, Diego A.},
  title     = {Modeling the Hidden Flexibility of Clustered Unit Commitment},
  journal   = {IEEE Transactions on Power Systems},
  volume    = {34},
  number    = {4},
  pages     = {3294--3296},
  year      = {2019},
  doi       = {10.1109/TPWRS.2019.2908051}
}

@article{AggregationLink,
  author    = {Meus, Jelle and Poncelet, Kris and Delarue, Erik},
  title     = {Applicability of a Clustered Unit Commitment Model in Power System Modeling},
  journal   = {IEEE Transactions on Power Systems},
  volume    = {33},
  number    = {2},
  pages     = {2195--2204},
  year      = {2018},
  doi       = {10.1109/TPWRS.2017.2736441}
}

@article{AggregationEF,
  author    = {Knueven, Ben and Ostrowski, Jim and Watson, Jean-Paul},
  title     = {Exploiting Identical Generators in Unit Commitment},
  journal   = {IEEE Transactions on Power Systems},
  volume    = {33},
  number    = {4},
  pages     = {4496--4507},
  year      = {2018},
  doi       = {10.1109/TPWRS.2017.2783850}
}

@article{AggregationTUA,
  author   = {Zhang, B. and Ding, T. and Xiao, Y.},
  title    = {A Tight Unit Aggregation for Unit Commitment to Eliminate Symmetry},
  journal  = {IEEE Transactions on Power Systems},
  volume   = {40},
  number   = {6},
  pages    = {5264--5275},
  month    = nov,
  year     = {2025},
  doi      = {10.1109/TPWRS.2025.3578915}
}

@article{AggregationTCUC,
  author   = {Najafi, Afshin and Pourmousavi, Mohammad A. and Amraee, Turaj},
  title    = {Improved Clustered Unit Commitment: Proposing Two Tight Models to Mitigate Infeasible Solution Space},
  journal  = {IEEE Transactions on Power Systems},
  year     = {2026},
  doi      = {10.1109/TPWRS.2026.3724406},
  note     = {Early access, doi: 10.1109/TPWRS.2026.3724406}
}

@misc{ReplicationCodeData,
  author = {Duan, Zixuan and Li, Zhengshuo},
  title  = {Online document and case studies},
  year   = {2026},
  note   = {[Online]. Available: \url{https://github.com/ZixuanDuan/CexAgg}}
}
\vspace{-6pt}
\end{document}